# TomoATT: An open-source package for Eikonal equation-based adjoint-state traveltime tomography for seismic velocity and azimuthal anisotropy

Jing Chen[a], Masaru Nagaso[a,1], Mijian Xu[a,2] and Ping Tong[a,b,c]

[a]Division of Mathematical Sciences, School of Physical and Mathematical Sciences, Nanyang Technological University, Singapore, Singapore

[b]Earth Observatory of Singapore, Nanyang Technological University, Singapore, Singapore

[c]Asian School of the Environment, Nanyang Technological University, Singapore, Singapore

Authorship contribution statement

**Jing Chen**: Writing the manuscript, writing the programs, developing the method. **Masaru Nagaso**: Refining the manuscript, establishing the program framework, writing the programs. **Mijian Xu**: Refining the manuscript, writing the programs. **Ping Tong**: Conceiving the project, refining the manuscript, developing the method.

ABSTRACT

---

[1] Masaru Nagaso is now at the Department of Applied Mathematics and Statistics, Colorado School of Mines, Golden, CO, USA.
[2] Mijian Xu is now at the Department of Physics, University of Toronto, Canada

TomoATT is an open-source software package, aiming at determining seismic velocity and azimuthal anisotropy based on adjoint-state traveltime tomography methods. Key features of TomoATT include Eikonal equation modeling, adjoint-state method, sensitivity kernel regularization, and multi-level parallelization. Through several toy experiments, we demonstrate TomoATT's capability in accurate forward modeling, handling multipathing phenomenon, delivering reliable tomographic results, and achieving high-performance parallelization. Additionally, TomoATT is benchmarked with a synthetic experiment and two real-data applications in central California near Parkfield and Thailand. The successful recovery of the synthetic model, along with the imaging results that are consistent with previous studies and regional tectonics, verifies the effectiveness of TomoATT. Each inversion starts with only three simple input files (about model, data, and parameters) and completes within 2 hours using 64 processors. Overall, TomoATT offers an efficient and user-friendly tool for regional and teleseismic traveltime tomography, empowering researchers to image subsurface structures and deepen our understanding of the Earth's interior.

## 1. Introduction

We developed an open-source software package, TomoATT, to invert traveltimes and differential arrival times from regional and teleseismic earthquakes for velocity heterogeneity and azimuthal anisotropy. This package aims to address the growing need for accurate subsurface seismic imaging, as well as to publicize the innovative adjoint-state traveltime tomography (ATT) methods (J. Chen, G. Chen, et al., 2023; J. Chen, S. Wu, et al., 2023; Tong, 2021a).

Compared with ray-based and wave equation-based tomography methods, the salient features of the ATT methods include the Eikonal equation-based forward modeling and the adjoint-state method for kernel calculation. Solving the Eikonal equation balances accuracy and efficiency: it is accurate and robust for calculating synthetic traveltime, avoiding the potential inaccuracy of bending and shooting methods for ray tracing (Rawlinson et al., 2008; Vidale, 1988). Additionally, its computational complexity is an order of magnitude lower than that of solving a wave equation. The adjoint-state method is applied to derive sensitivity kernels, which has proven effective in partially addressing the multipathing phenomenon (Tong et al., 2023), thus yielding accurate sensitivity kernels. During the inversion procedure, the step size-controlled gradient descent method is employed to update model parameters, with the multiple-grid parameterization and kernel density regularization incorporated to enhance the inversion reliability and accelerate convergence. These attributes make the ATT methods an efficient and robust tomographic approach for imaging subsurface velocity heterogeneity and seismic anisotropy (J. Chen, G. Chen, et al., 2023; J. Chen, S. Wu, et al., 2023; Tong, 2021a, 2021b; Wu et al., 2022).

The TomoATT package is developed in C++ based on the ATT methods. It utilizes the Message Passing Interface (MPI) and shared memory techniques for efficient parallelization. Popular file formats are supported for input and output operations, including YAML for parameter file, HDF5 for model file, and TEXT for data file. To improve user-friendliness, we also develop PyTomoATT, a companion Python module to streamline the processing of input and output files. All resources are accessible through the link provided in the Code Availability Section.

This paper provides a comprehensive introduction to TomoATT. In Section 2, we briefly review the ATT methods and present an overview of the package. Section 3 demonstrates the key features of TomoATT through numerical experiments. Furthermore, we evaluate the performance of TomoATT using both synthetic and real data in Section 4, showcasing its capability to resolve complex subsurface structures. By sharing our experiences and insights, we hope to encourage the geophysical community to adopt and further develop this open-source tool, advancing the field of seismic imaging and improving our understanding of subsurface dynamics.

## 2. Methodology and overview of TomoATT

### 2.1 Brief review of the ATT methods

Here we briefly review the ATT methods (G. Chen et al., 2023; J. Chen, G. Chen, et al., 2023; Tong, 2021a; Tong et al., 2023). These methods aim at the determination of seismic slowness $s(\boldsymbol{x})$ (the reciprocal of velocity) and azimuthal anisotropy parameters $\xi(\boldsymbol{x})$ and $\eta(\boldsymbol{x})$ by minimizing the discrepancy between synthetic and observational traveltimes, common-source differential arrival times, and common-receiver differential arrival times. The related problem is formulated as the following optimization problem

$$\min_{s,\xi,\eta} \chi = \alpha\chi_t + \beta\chi_{cs} + \gamma\chi_{cr}. \tag{1}$$

The objective function is a weighted sum of the misfit functions corresponding to traveltime, common-source differential arrival times, and common-receiver differential arrival times, respectively, given by

$$\chi_t = \sum_{n=1}^{N_s}\sum_{m=1}^{N_r} \frac{w_{n|m}}{2}\left(T_n(\boldsymbol{x}_{r,m}) - T_n^{obs}(\boldsymbol{x}_{r,m})\right)^2, \tag{2}$$

$$\chi_{cs} = \sum_{n=1}^{N_s} \sum_{m=1}^{N_r} \sum_{i=1}^{N_r} \frac{w_{n|m,i}}{2} \left( \left( T_n(x_{r,m}) - T_n(x_{r,i}) \right) - \left( T_n^{obs}(x_{r,m}) - T_n^{obs}(x_{r,i}) \right) \right)^2, \quad (3)$$

$$\chi_{cr} = \sum_{n=1}^{N_s} \sum_{j=1}^{N_s} \sum_{m=1}^{N_r} \frac{w_{n,j|m}}{2} \left( \left( T_n(x_{r,m}) - T_j(x_{r,m}) \right) - \left( T_n^{obs}(x_{r,m}) - T_j^{obs}(x_{r,m}) \right) \right)^2. \quad (4)$$

$\alpha, \beta, \gamma$ are weights for three types of data. $T_n(x_{r,m})$ and $T_n^{obs}(x_{r,m})$ represent the synthetic and observational traveltimes of a particular seismic phase emanating from the $n$-th source and recorded by the $m$-th receiver, respectively. The weight coefficients $w_{n|m}, w_{n|m,i}, w_{n,j|m}$ are determined by the quality of the observational data.

The ATT method derives the sensitivity kernels of the objective function with respect to model parameters based on the adjoint-state method and solves the optimization problem iteratively (Figure 1). Each iteration consists of four key steps:

1. Calculate the synthetic traveltime field $T_n(x)$ for each source $x_{s,n}$ by solving the anisotropic Eikonal equation in spherical coordinates

$$[\nabla T_n(x)]^t M(x;\xi,\eta) \nabla T_n(x) = \left( \partial_r T_n \quad \frac{\partial_\theta T_n}{r} \quad \frac{\partial_\phi T_n}{r\cos\theta} \right) \begin{pmatrix} 1 & 0 & 0 \\ 0 & 1-2\xi & 2\eta \\ 0 & 2\eta & 1+2\xi \end{pmatrix} \begin{pmatrix} \partial_r T_n \\ \frac{\partial_\theta T_n}{r} \\ \frac{\partial_\phi T_n}{r\cos\theta} \end{pmatrix} = s^2(x). \quad (4)$$

The boundary condition is set as $T_n(x_{s,n}) = 0$ for earthquakes within the study region $\Omega$ or, for teleseismic earthquakes, is set as the travel time from the source to the study region boundary $\partial\Omega$, by solving 2D Eikonal equations (Chen, Wu, et al., 2023).

2. Calculate the adjoint field $P_n(x)$ based on each traveltime field $T_n(x)$ by solving the adjoint equation

$$\nabla \cdot \left( P_n(x) ([-\nabla T_n(x)]^t M(x;\xi,\eta)) \right) = \sum_{m=1}^{N_r} R_{n,m} \delta(x - x_{r,m}). \quad (5)$$

Here the adjoint source takes the general form of $R_{n,m} = \frac{\partial \chi}{\partial T_n(x_{r,m})}$, which is specified in Appendix A.

3. Calculate the sensitivity kernels with respect to slowness $s(x)$ and anisotropic parameters $\xi(x), \eta(x)$ based on the adjoint-state method

$$K_s(x) = \sum_{n=1}^{N_s} P_n(x) s^2(x), \quad (6)$$

$$K_\xi(x) = \sum_{n=1}^{N_s} P_n(x) \left( \left( \frac{\partial_\theta T_n(x)}{r} \right)^2 - \left( \frac{\partial_\phi T_n(x)}{r \cos\theta} \right)^2 \right), \tag{7}$$

$$K_\eta(x) = \sum_{n=1}^{N_s} -2P_n(x) \frac{\partial_\theta T_n(x)}{r} \frac{\partial_\phi T_n(x)}{r \cos\theta}. \tag{8}$$

An approximate linear relationship between the perturbations of model parameters and objective function is accordingly constructed as follows

$$\delta\chi = \int_\Omega K_s(x) \frac{\delta s(x)}{s(x)} dx + \int_\Omega K_\xi(x) \delta\xi(x) dx + \int_\Omega K_\eta(x) \delta\eta(x) dx. \tag{9}$$

4. Update $s(x), \xi(x), \eta(x)$ using the step size-controlled gradient descent method (Chen, Chen, et al., 2023). The multiple-grid parameterization and kernel density regularization techniques are applied to enhance reliability.

The influence of source uncertainty on imaging results can be mitigated by updating hypocenter $x_{s,n}$ and origin time $\tau_n$ during the inversion using the gradient descent method. The sensitivity kernels of the objective function with respect to $x_{s,n}$ and $\tau_n$ take the general forms of

$$K_{x_{s,n}} = \sum_{m=1}^{N_r} \frac{\partial \chi}{\partial \Gamma_m(x_{s,n})} \nabla \Gamma_m(x_{s,n}) \text{ and } K_{\tau_n} = \frac{\partial \chi}{\partial \tau_n}, \tag{10}$$

which are specified in Appendix B.

**2.2 Overview of the TomoATT package**

One of the core features of the package is its robust utilization of the MPI, which allows for effective multi-level parallelization. This design empowers users to leverage high-performance computing (HPC) platforms, maximizing computational resource efficiency, which is particularly beneficial for large-scale seismic imaging applications. Additionally, we employ shared memory techniques to optimize memory usage and enhance parallel processing efficiency.

To improve accessibility, TomoATT supports popular file formats for input and output operations. The package accommodates the YAML format for parameter files, which provides an intuitive way for users to configure their applications. For storing model parameters, the parallel version of HDF5 is utilized, renowned for its ability to handle large dataset efficiently. This format allows multiple processes to read from and write to data files simultaneously, enhancing data access speed and overall computational efficiency. Additionally, a TEXT format is supported for managing input traveltime data, often characterized by a non-regular structure. To address the challenges posed by this structure, we developed PyTomoATT, a companion Python module designed to process and handle non-regular data effectively. PyTomoATT includes functionalities for data filtering, weight assignment, coordinate rotation, and other preprocessing tasks. Moreover, it provides tools for model generation and reading, model slicing, checkerboard generation, and the creation and modification of parameter files. These capabilities work together to streamline the preparation and management of input and output files, enabling researchers to set up the inversion process efficiently without needing extensive programming knowledge.

All resources related to TomoATT, including the core package, user documentation, and PyTomoATT, are licensed under the GNU General Public License v3.0. These tools are compatible with multiple platforms, including MacOS, Linux, and Windows through Windows Subsystem for Linux (WSL). TomoATT can be compiled with various compilers, including GNU, Intel, and Clang, ensuring broad accessibility for users across different operating systems and development environments. The tools and documentation are available through the link provided in the Code availability section.

## 3. Key features of TomoATT

In this section, we design several toy experiments to demonstrate the key features of TomoATT, including forward modeling, adjoint-state method, multiple-grid parameterization, kernel density regularization, and multi-level parallelization.

### 3.1 Forward modeling

TomoATT solves the anisotropic Eikonal equation for the traveltime field of wavefront propagation from the source to any positions within the study region. We select the fast sweeping method (FSM) (Zhao, 2005) as the Eikonal solver. This grid-based method has proven unconditionally convergent to the solution (Zhao, 2005), and achieves the optimal computational complexity of $O(N)$, where $N$ is the total number of grid nodes. Additionally, we use the multiplicative factorization technique (Luo & Qian, 2012) to eliminate source singularity and solve the equation in spherical coordinates to account for Earth's curvature (Chen, Chen, et al., 2023), further improving accuracy. Here we design two toy models to evaluate the accuracy of FSM for calculating traveltime and teleseismic differential arrival time, respectively.

First, we consider a linear velocity model within the domain ranging from $[0, 20°] \times [0, 400 \text{ km}]$, where the velocity approximately increases from 4.5 km/s at the surface to 9 km/s at the bottom (Figure 2a). The earthquake is located at the center $x_s = (10°, 200 \text{ km})$. The Eikonal equation is solved on four meshes with grid spacings of 20, 10, 5, and 2.5 km, respectively. The $L_1$ norm error between the calculated and analytical solutions $\|T_{cal}(x) - T_{true}(x)\|_{L_1}$ decreases as grid spacing decreases, from 0.0517 s to 0.0096 s (Figure 2c), demonstrating the expected first-order accuracy (Luo & Qian, 2012). We also measure the traveltime errors at stations on the surface, which accumulate with epicenter distance (Figure 2b). The results show that the maximum error in this model remains below 0.1 s for epicenter distances less than 800 km, provided grid spacing is less than 5 km.

Second, we consider the global AK135 model (Kennett et al., 1995) within the region $[11°N, 21°N] \times [96°E, 106°E] \times [0, 500 \text{ km}]$. A teleseismic earthquake is located at $(5°S, 142°E, 20 \text{ km})$, with an epicenter distance of 45.68° from the center of the study region (Figure 3a). 81 stations are evenly distributed on the surface (Figure 3b). We solve the Eikonal equation on a mesh with grid spacing of 10 km × 10 km × 5 km, and calculate common-source differential traveltimes between stations less than 300 km apart. Since the analytical solution is unavailable, we compare results obtained from the Eikonal solver with those from the TauP software (Crotwell et al., 1999). The standard deviation of misfits $(\Delta T_{cal} - \Delta T_{TauP})$ is 0.03 s (Figure 3c), verifying the accuracy of our Eikonal solver.

### 3.2 Sensitivity kernels based on the adjoint-state method

TomoATT computes sensitivity kernels using the adjoint-state method. It solves for adjoint field in the study region, which describes the transportation of adjoint sources from receivers to the source along the opposite direction of wavefront propagation (e.g., along the negative direction of traveltime gradient in isotropic media). The main advantage is that this approach measures the sensitivity field throughout the study region, rather than restricting it to a single ray path. Therefore, it effectively addresses multipathing phenomena, where the wavefront reaches one receiver along multiple paths with similar traveltimes (Tong et al., 2023).

Here we present an example to illustrate this issue. We consider a linear velocity model with several high- and low-velocity anomalies embedded (Figure 4a). An earthquake is located at 40 km depth, and three stations are deployed on the surface. Two paths exist from the source to each station with similar minimum traveltimes. Consequently, the sensitivity kernel should be supported on both paths, as the velocity perturbation along either path would affect the traveltime. The adjoint-state method successfully provides correct sensitivity kernels by measuring the sensitivity throughout the study region (Figure 4b). In contrast, without special treatment, ray-based methods identify only a single ray path, resulting in an incomplete sensitivity kernel.

**3.3 Multiple-grid parameterization and kernel density normalization**

After obtaining the sensitivity kernels on discretized grid nodes, TomoATT applies two regularization techniques to enhance inversion reliability: multiple-grid parameterization and kernel density normalization. The grid nodes for forward modeling are typically dense to ensure the accuracy. Directly updating model parameters on the forward grid nodes is unsuitable, as the number of variables may exceed the resolving ability of the limited traveltime data. The multiple-grid parameterization (Tong et al., 2019) has been proposed to address this problem.

The multiple-grid parameterization method designs $H$ sets of coarse inversion grids, denoting the nodes of the $h$-th grid as $(r_i^h, \theta_j^h, \phi_k^h)$. A series of basis functions are defined based on these inversion grids, given by

$$B_l^h(r, \theta, \phi) = \frac{1}{H} u_i^h(r) v_j^h(\theta) w_k^h(\phi), \quad l = i + N_I^h(j-1) + N_I^h N_J^h(k-1), \tag{11}$$

in which

$$u_i^h(r) = \begin{cases} (r - r_{i-1}^h)/(r_i^h - r_{i-1}^h), & r_1^h \leq r_{i-1}^h \leq r \leq r_i^h, \\ (r_{i+1}^h - r)/(r_{i+1}^h - r_i^h), & r_i^h \leq r \leq r_{i+1}^h \leq r_{N_I}^h, \\ 0, & \text{otherwise}. \end{cases} \tag{12}$$

$$v_i^h(\theta) = \begin{cases} (\theta - \theta_{j-1}^h)/(\theta_j^h - \theta_{j-1}^h), & \theta_1^h \leq \theta_{j-1}^h \leq \theta \leq \theta_j^h, \\ (\theta_{j+1}^h - \theta)/(\theta_{j+1}^h - \theta_j^h), & \theta_j^h \leq \theta \leq \theta_{j+1}^h \leq \theta_{N_J}^h, \\ 0, & \text{otherwise}. \end{cases} \tag{13}$$

$$w_i^h(\phi) = \begin{cases} (\phi - \phi_{k-1}^h)/(\phi_k^h - \phi_{k-1}^h), & \phi_1^h \leq \phi_{k-1}^h \leq \phi \leq \phi_k^h, \\ (\phi_{k+1}^h - \phi)/(\phi_{k+1}^h - \phi_k^h), & \phi_k^h \leq \phi \leq \phi_{k+1}^h \leq \phi_{N_K}^h, \\ 0, & \text{otherwise}. \end{cases} \tag{14}$$

A model perturbation field is then assumed to be represented by a linear combination of these basis functions

$$\delta m = \sum_{l,h} \delta C_l^h B_l^h(\mathbf{x}). \tag{15}$$

Consequently, the approximate linear relationship (9) between the perturbations in model parameters and the objective functions is modified to

$$\delta \chi = \sum_{l,h} \frac{\partial \chi}{\partial C_{l,s}^h} \delta C_{l,s}^h + \sum_{l,h} \frac{\partial \chi}{\partial C_{l,\xi}^h} \delta C_{l,\xi}^h + \sum_{l,h} \frac{\partial \chi}{\partial C_{l,\eta}^h} \delta C_{l,\eta}^h, \tag{16}$$

where the new sensitivity kernel with respect to auxiliary parameters $\left(C_{l,s}^h, C_{l,\xi}^h, C_{l,\eta}^h\right)$ are

$$\frac{\partial \chi}{\partial C_{l,s}} = \int_\Omega K_s(\boldsymbol{x}) B_l^h(\boldsymbol{x}) \mathrm{d}\boldsymbol{x}, \quad \frac{\partial \chi}{\partial C_{l,\xi}} = \int_\Omega K_\xi(\boldsymbol{x}) B_l^h(\boldsymbol{x}) \mathrm{d}\boldsymbol{x}, \quad \frac{\partial \chi}{\partial C_{l,\eta}} = \int_\Omega K_\eta(\boldsymbol{x}) B_l^h(\boldsymbol{x}) \mathrm{d}\boldsymbol{x}. \tag{17}$$

This method projects the model perturbation $\delta s(\boldsymbol{x}), \delta \xi(\boldsymbol{x}), \delta \eta(\boldsymbol{x})$ from an infinite-dimensional function space into a finite-dimensional vector space represented by $\left(C_{l,s}^h, C_{l,\xi}^h, C_{l,\eta}^h\right)$. Thus, the number of variables is decreased, and the sensitivity kernel is smoothed by convoluting it with the basis function $B_l^h(\boldsymbol{x})$. Additionally, averaging contributions across multiple inversion grids mitigates potential error from subjective selection of a single inversion grid, thereby, enhancing inversion stability.

Real-data inversions often suffer from uneven data distribution, leading to slower model updates in regions with sparse data coverage. Various methods have been proposed to address this issue, such as event declustering (Tong, 2021b) and assigning weights (Ruan et al., 2019). Here, we consider the characteristics of the adjoint-state method and propose an innovative approach. We observe that the density of an event sensitivity kernel can be represented as a specific adjoint field $\widehat{P}_n(\boldsymbol{x})$, which follows the adjoint equation (5) with adjoint sources $R_{n,m} = 1$. This field can be used to normalize the sensitivity kernel by

$$K_s(\boldsymbol{x}) \leftarrow \frac{K_s(\boldsymbol{x})}{(K_d(\boldsymbol{x}) + \epsilon)^\zeta}, \quad K_\xi(\boldsymbol{x}) \leftarrow \frac{K_\xi(\boldsymbol{x})}{(K_d(\boldsymbol{x}) + \epsilon)^\zeta}, \quad K_\eta(\boldsymbol{x}) \leftarrow \frac{K_\eta(\boldsymbol{x})}{(K_d(\boldsymbol{x}) + \epsilon)^\zeta}, \tag{18}$$

where the kernel density $K_d(\boldsymbol{x})$ is formulated as

$$K_d(\boldsymbol{x}) = \sum_{n=1}^{N_s} \widehat{P}_n(\boldsymbol{x}). \tag{19}$$

The small value $\epsilon$ avoids division by zero, and the coefficient $\zeta$ controls the degree of normalization. The normalization emphasizes the contribution of data in sparsely constrained regions, which accelerates the model update in these areas. However, it also magnifies the influence of noise on imaging results. Thus, an appropriate coefficient $\zeta$ should be carefully determined based on data distribution and noise level.

Here we present a toy experiment to illustrate the multiple-grid parameterization and kernel density regularization. The study region is defined as $[0\ \mathrm{km}, 220\ \mathrm{km}] \times [0\ \mathrm{km}, 50\ \mathrm{km}]$, with 8 stations evenly deployed on the surface. To mimic an uneven earthquake distribution, we randomly place 10,000 earthquakes, with a higher probability in the upper left (Figure 5a). The initial model is an isotropic model where velocity increases from 4 km/s at the surface to 8 km/s at 40 km depth. The target model is designed as a checkerboard model by adding staggered velocity perturbations to the initial model (Figure 5b). The observational data is generated by calculating synthetic traveltimes in the target model and adding random Gaussian noise with a standard derivation of 0.1 s. We start with the initial model and invert the observational data to examine the recovery of checkerboard anomalies.

First, we perform the inversion without applying multiple-grid parameterization and kernel density normalization. The sensitivity kernel at the first iteration and the final inversion result are displayed in Figure 5c-5d. Numerous small artificial anomalies are observable at 1 km, and the geometry of anomalies above 20 km depth is distorted. These likely arise from the excessive number of variables leading to unreliable inversion results. Second, we apply the multiple-grid parameterization, which eliminates the artifacts and produces a more reasonable inversion result (Figure 5e-5f). However, due to uneven earthquake distribution, the anomalies at the right bottom are less well recovered. Furthermore, we apply the kernel density normalization with different coefficients $\zeta = 0.3, 0.6, 0.9$ (Figure 5g-5l). The normalization indeed emphasizes the sensitivity kernel at the right bottom, leading to better recovery of anomalies and reducing smearing caused by uneven earthquake distribution. A higher value of $\zeta$ leads to a higher degree of anomaly recovery; however, it also magnifies the artificial anomalies below 40 km depth at the right bottom due to insufficient data constraints.

### 3.4 Multi-level parallelization

In TomoATT, solving anisotropic Eikonal equations and adjoint equations at each iteration constitutes the primary computation cost. To fully utilize computational resources on high-performance-computing platforms, TomoATT implements three levels of parallelization (Figure 6a): Source parallelization, domain decomposition (Zhao, 2007), and hyperplane stepping parallelization (Detrixhe & Gibou, 2016; Detrixhe et al., 2013).

Source parallelization (Figure 6a, Level 1) is the most straightforward approach, as solving Eikonal and adjoint equations corresponding to multiple sources are completely independent. Consequently, source parallelization can achieve high parallel efficiency. The primary drawback is the significant memory required for simultaneously solving three-dimensional (3-D) equations, which may impose a considerable memory burden.

Domain decomposition is a widely applicable parallelization approach, which divides the study region into multiple subdomains and assigns one processor to solve equations within each subdomain (Figure 6a, Level 2). Its main advantage is that it requests almost no additional memory. However, a drawback is the increased computational cost due to boundary communications between processors managing adjacent subdomains. More importantly, increasing the number of subdomains may result in more iterations for the FSM to converge, decreasing parallel efficiency.

Additionally, hyperplane stepping parallelization enables multiple processors to solve equations within a single subdomain (Figure 6a, Level 3). In this method, $N_x \times N_y \times N_z$ 3-D grid nodes are categorized into $N_x + N_y + N_z - 2$ hyperplanes. Handling multiple grid nodes within the same hyperplane is independent, thus allowing for parallelized computation. This approach, like domain decomposition, requires no additional memory. By leveraging a shared memory technique in MPI, communication between processors is avoided, which helps achieve good parallel efficiency.

We evaluate the performance of three parallelization methods at the National Supercomputing Centre (NSCC) Singapore, using up to 32 processors to solve 64 Eikonal equations on a mesh of $101 \times 101 \times 101$ grid nodes. The speed-up and memory usage with respect to the number of processors are displayed in Figure 6b and Table 1. Source parallelization achieves the highest parallel efficiency but requires substantial memory. In comparison, domain decomposition and hyperplane stepping parallelization almost require no additional memory, though they yield slightly lower parallel efficiency. In summary, if memory capacity is sufficient, prioritizing processors for source parallelization is optimal, followed by hyperplane stepping parallelization and domain decomposition.

## 4 Case studies

In this section, we first validate TomoATT using a synthetic experiment. Subsequently, we benchmark the package with two real-data inversion cases: regional tomography in California near Parkfield, and teleseismic tomography in Thailand and adjacent areas. The tomography for these two real cases has been detailed in our previous studies (Chen, Chen, et al., 2023; Chen, Wu, et al., 2023).

### 4.1 Earthquake location and tomography of a synthetic model

We design such a synthetic experiment, merely aiming at verifying earthquake location and tomography functions in TomoATT. The study region is $[0, 2°] \times [0, 2°] \times [0, 40 \text{ km}]$, with 25 stations uniformly deployed on the surface. The true model is built by embedding staggered velocity and anisotropic perturbations to an isotropic background model, whose velocity linearly increases from 5.0 km/s at 0 km depth to 8.0 km/s at 40 km depth. 867 earthquakes are regularly located at depths of 10, 20, and 30 km (Figure 7a). Observational traveltime data are generated using the true model and true earthquake locations. Correspondingly, the initial model is the background velocity model without any velocity or anisotropic perturbations. The initial earthquake locations are randomly deviated from the true locations, following uniform distributions with standard deviations of 0.1° in latitude and longitude, 10 km in depth, and 0.5 s in origin time (Figure 7b)

We perform the following 5 tests:

- Test 1: Locate earthquakes in the true model. The inversion starts with the initial earthquake locations and the true model, performing 200 iterations to update earthquake locations while keeping model parameters fixed.
- Test 2: Locate earthquakes in the initial model. The procedure is the same as in Test 1 but uses the initial model instead.
- Test 3: Update model parameters using the true earthquake locations. The inversion starts with the true earthquake location and the initial model, performing 40 iterations to update velocity and anisotropic parameters while keeping earthquake locations fixed.
- Test 4: Update model parameters using the initial earthquake locations. The procedure is the same as in Test 2 but uses the initial earthquake locations instead.
- Test 5: Simultaneously locate earthquakes and update model parameters. The inversion is conducted in three stages to reduce variable coupling. First, we perform a preliminary earthquake location in the initial model for 50 iterations. Second, we simultaneously update model parameters and earthquake locations for 40 iterations. Finally, we relocate earthquakes in the improved velocity model for 100 iterations to achieve more accurate earthquake locations.

In Test 1, earthquakes are accurately located in the true model (Figure 8a), verifying the relocation function in TomoATT. In contrast, Test 2 shows that final earthquake locations are slightly offset when using inaccurate velocity and anisotropic model parameters (Figure 8b), highlighting the importance of an accurate model for precise earthquake location. Similarly, Test 3 indicates that the staggered velocity and anisotropic perturbations are accurately imaged using the true earthquake locations (Figure 8c), verifying the imaging function in TomoATT. However, in Test 4, inaccurate earthquake locations result in significant distortions and artifacts in the imaged velocity and anisotropy (Figure 8d), which underlines the critical role of accurate earthquake location in tomography. In practical applications, both initial earthquake locations and model parameters are typically biased. Therefore, it is suggested to update both earthquake locations and model parameters by simultaneous inversion, as illustrated in Test 5 (Figure 8e). This test result indicates that both model parameters and earthquake locations can be effectively constrained using the simultaneous inversion function in TomoATT.

### 4.2 Regional tomography in central California near Parkfield

We benchmark TomoATT with the regional tomography in central California near Parkfield, covering a study region of 160 km×440 km horizontally and 50 km vertically (Figure 9a). After applying strict data selection criteria (see more details in Chen, Chen, et al. (2023)) to the traveltime data from Northern California Earthquake Data Center (NCEDC, 2014) and Southern California Earthquake Data Center (SCEDC, 2013), we collected 1,218,044 first P-arrival times from 32,721 earthquakes and recorded by 607 stations for inversion. As the earthquake locations has been robustly determined by the data centers, we only inverted for velocity heterogeneity and azimuthal anisotropy, keeping earthquake locations fixed.

The study region is discretized into a mesh of 149,787 grid nodes, with an approximate grid spacing of **5 km × 5 km × 1 km**. This grid spacing is sufficient to ensure the accuracy of our Eikonal solver, as validated in Section 3.1. By applying the reciprocal principle, we can regard the 607 stations as sources and the 32,721 earthquakes as receivers, which allows for numerically solving only 607 anisotropic Eikonal equations and adjoint equations per iteration, rather than 32,721. We perform 80 iterations on the NSCC platform using 64 processors powered by AMD EPYC 7713 64-Core Processor of 2.0 GHz, completing the inversion in 57 minutes.

Pronounced velocity perturbations are revealed in the imaging results, illustrated in horizontal sections in Figure 9b. At 4 km depth, the San Andreas Fault (SAF) clearly separates the low-velocity zone beneath the Franciscan terrane (FT) in the west from the high-velocity anomaly beneath the Salinian terrane (ST) in the east. This characteristic structure aligns well with previous imaging results (Eberhart-Phillips & Michael, 1993; Lippoldt et al., 2017; Thurber et al., 2006). The velocity contrast remains at 8 km depth, though the high-velocity anomaly weakens. By 16 km depth, it is replaced by a broad low-velocity anomaly beneath the SAF. This anomaly is suggested by ambient noise tomography (Lippoldt et al., 2017) to expand to the lower crust and uppermost mantle. It is also associated with a high-conductivity zone in the crust (Becken et al., 2011), possibly indicating the presence of crustal fluid (Tong, 2021). Our result also images a low-velocity anomaly beneath the Santa Maria Basin (SMB), which extends from the surface to 8 km and is connected to the broad low-velocity anomaly at 16 km.

The crustal azimuthal anisotropy beneath the SAF and vicinity shows distinct patterns (Figure 9c). Specifically, strong anisotropy is observed beneath the creeping and transition segments of the SAF at 8 km depth, with the fast velocity direction parallel to the fault. This suggests that anisotropy near the SAF is dominated by the fault structure (Boness & Zoback, 2006). This observation is consistent with the results of teleseismic receiver function analysis (Audet, 2015; Ozacar & Zandt, 2009). In contrast, farther from the SAF, the fast velocity direction shifts, forming a large angle or becoming nearly perpendicular to the fault. This pattern coincides with the maximum horizontal compression ($S_{Hmax}$) direction observed from borehole data (Townend & Zoback, 2004), indicating that the anisotropy is stress-induced.

In summary, variations in seismic velocity and azimuthal anisotropy in central California near Parkfield are successfully imaged by inverting first P-arrival traveltimes. The consistency of the results with local tectonics and previous studies validates the effectiveness of TomoATT in the regional tomography.

### 4.3 Teleseismic tomography in Thailand and adjacent areas

We benchmark TomoATT with the teleseismic tomography in Thailand and adjacent areas. A total of 51 stations, consisting of 40 temporary stations from the Thai Seismic Array (Tanaka et al., 2019) and 11 additional permanent stations, are deployed within the study region spanning from **97°E** to **106°E** and from **12°N** to **21°N** (Figure 10a). We download waveforms from 190 earthquakes with epicentral distances between **30°** and **90°** (Figure 10b) and use the multi-channel cross correlation technique (VanDecar & Crosson, 1990) to extract common-source differential arrival times. After applying strict data selection criteria (Chen et al., 2023), we collect 15,205 common-source differential arrival times to invert for upper mantle seismic velocity. The azimuthal anisotropy is fixed as 0 during the inversion, as the nearly vertical ray paths of teleseismic events provide limited constraints on azimuthal anisotropy.

The computational region of **12° × 12° × 600 km** is discretized into a mesh of 1,771,561 grid nodes, with an approximate grid spacing of **10 km × 10 km × 5 km**. This grid spacing is sufficient to ensure the accuracy of our Eikonal solver, as validated in Section 3.2. We perform 80 iterations on the NSCC platform using 64 processors, completing the inversion in 124 minutes.

Figure 10c illustrates the horizontal sections of velocity perturbations related to the horizontal average. The most notable feature is the high-velocity perturbation beneath the Khorat Plateau, contrasting with a low-velocity perturbation to the west. This structure shows consistency with previous tomographic images (Li et al., 2006; Yang et al., 2015), and corresponds to a cold and thick lithosphere beneath the Khorat Plateau, as inferred by the thermal analysis based on S-velocity (C. Yu et al., 2017). Furthermore, our result reveals a low-velocity perturbation beneath the southwestern corner of the Khorat Plateau, different from the high-velocity perturbation in the central area. This distinct feature may suggest that the southwestern margin of the Khorat Plateau is partially modified. The vertical section BB^' crossing this anomaly reveals two low-velocity channels extending from the surface down to the upper mantle. These low-velocity perturbations may indicate the pathways for mantle material upwelling, potentially driven by the mantle convection associated with surrounding subduction systems of the Indo-Australian, Pacific, and Philippine Sea Plates (Lin et al., 2019). Notably, the western channel appears connected to a possible slab window of the Indian Plate (Pesicek et al., 2008; Y. Yu et al., 2017). The mantle upwelling through the slab window may also be a contributing factor to the mantle upwelling (Arboit et al., 2016).

## 5 Conclusions

TomoATT provides an effective solution for regional and teleseismic traveltime tomography, based on the Eikonal equation-based adjoint-state traveltime tomography methods. Synthetic experiments showcase its advantages in accurate forward modeling, handling multipathing phenomenon, sensitivity kernel regularization for reliable inversion, and multi-level parallelization for high efficiency. Users can initiate an inversion easily with only three simple files: an HDF5 model file, a YAML parameter file, and a TEXT data file. Two real-data applications—regional tomography in central California near Parkfield and teleseismic tomography in Thailand and adjacent regions—were completed within 2 hours using 64 processors. The imaging results reveal pronounced velocity perturbations and azimuthal anisotropies that align well with previous studies and local tectonics. These benchmarks demonstrate the effectiveness and efficiency of TomoATT for regional and teleseismic tomography.

## 6 Acknowledgments

This work was supported by the Minister of Education, Singapore, under its MOE AcRF Tier-2 Grant (MOE-T2EP20122-0008). The computational work for this article was partially performed on resources of the National Supercomputing Centre, Singapore (https://www.nscc.sg). The authors acknowledge the seismic data for benchmark, available from the Northern California Earthquake Data Center (NCEDC 2014), the Southern California Earthquake Data Center (SCEDC 2013), TSAR (Tanaka et al., 2019) (http://ohpdmc.eri.u-tokyo.ac.jp/breq-fast-tsar/index.html),

and the Incorporated Research Institutions for Seismology (IRIS) Data Management Center (http://ds.iris.edu/mda/TM/PBKT/?starttime=2008-01-01&endtime=2599-12-31).

**Code availability section**

Name of the code/library: TomoATT, PyTomoATT

License: GNU General Public License v3.0

Contact: tongping@ntu.edu.sg, jing.chen@ntu.edu.sg

Hardware requirements: PC or HPC

Program language: C++, Python

Software required: C++ compiler, CMake, OpenMPI/MPICH, HDF5, YAML, Python 3.9 or later

Program size: 9.53 MB

The source codes and document are available at https://tomoatt.com/

# References

Arboit, F., Collins, A. S., Morley, C. K., Jourdan, F., King, R., Foden, J., Amrouch, K., 2016. Geochronological and geochemical studies of mafic and intermediate dykes from the Khao Khwang Fold–Thrust Belt: implications for petrogenesis and tectonic evolution. Gondwana Research 36, 124-141.

Audet, P., 2015. Layered crustal anisotropy around the San Andreas Fault near Parkfield, California. Journal of Geophysical Research: Solid Earth 120, 3527-3543.

Becken, M., Ritter, O., Bedrosian, P. A., Weckmann, U., 2011. Correlation between deep fluids, tremor and creep along the central San Andreas fault. Nature 480, 87-90.

Boness, N. L., Zoback, M. D., 2006. Mapping stress and structurally controlled crustal shear velocity anisotropy in California. Geology 34, 825-828.

Chen, G., Chen, J., Tape, C., Wu, H., Tong, P., 2023. Double-difference adjoint tomography of the crust and uppermost mantle beneath Alaska. Journal of Geophysical Research: Solid Earth 128, e2022JB025168.

Chen, J., Chen, G., Nagaso, M., Tong, P., 2023. Adjoint-state traveltime tomography for azimuthally anisotropic media in spherical coordinates. Geophysical Journal International 234, 712-736.

Chen, J., Wu, S., Xu, M., Nagaso, M., Yao, J., Wang, K., Li, T., Bai, Y., Tong, P., 2023. Adjoint-state teleseismic traveltime tomography: method and application to Thailand in Indochina Peninsula. Journal of Geophysical Research: Solid Earth 128, e2023JB027348.

Crotwell, H. P., Owens, T. J., Ritsema, J., 1999. The TauP Toolkit: Flexible seismic travel-time and ray-path utilities. Seismological Research Letters 70, 154-160.

Detrixhe, M., Gibou, F., Min, C., 2013. A parallel fast sweeping method for the Eikonal equation. Journal of Computational Physics 237, 46-55.

Detrixhe, M., Gibou, F., 2016. Hybrid massively parallel fast sweeping method for static Hamilton–Jacobi equations. Journal of Computational Physics 322, 199-223.

Eberhart-Phillips, D., Michael, A. J., 1993. Three-dimensional velocity structure, seismicity, and fault structure in the Parkfield region, central California. Journal of Geophysical Research: Solid Earth 98, 15737-15758.

Kennett, B. L. N., Engdahl, E. R., Buland, R., 1995. Constraints on seismic velocities in the Earth from traveltimes. Geophysical Journal International 122, 108-124.

Li, C., van der Hilst, R. D., Toksöz, M. N., 2006. Constraining P-wave velocity variations in the upper mantle beneath Southeast Asia. Physics of the Earth and Planetary Interiors 154, 180-195.

Lin, J., Xu, Y., Sun, Z., Zhou, Z., 2019. Mantle upwelling beneath the South China Sea and links to surrounding subduction systems. National Science Review 6, 877-881.

Lippoldt, R., Porritt, R. W., Sammis, C. G., 2017. Relating seismicity to the velocity structure of the San Andreas Fault near Parkfield, CA. Geophysical Journal International, 209, 1740-1745.

Luo, S., Qian, J., 2012. Fast sweeping methods for factored anisotropic Eikonal equations: multiplicative and additive factors. Journal of Scientific Computing 52, 360-382.

NCEDC, 2014. Northern California Earthquake Data Center. UC Berkeley Seismological Laboratory, Dataset, doi:10.7932/NCEDC.

Ozacar, A. A., Zandt, G., 2009. Crustal structure and seismic anisotropy near the San Andreas Fault at Parkfield, California. Geophysical Journal International, 178, 1098-1104.

Pesicek, J., Thurber, C., Widiyantoro, S., Engdahl, E., DeShon, H., 2008. Complex slab subduction beneath northern Sumatra. Geophysical Research Letters 35, L20303.

Rawlinson, N., Hauser, J., Sambridge, M., 2008. Seismic ray tracing and wavefront tracking in laterally heterogeneous media. Advances in geophysics 49, 203-273.

Ruan, Y., Lei, W., Modrak, R., Örsvuran, R., Bozdağ, E., Tromp, J., 2019. Balancing unevenly distributed data in seismic tomography: a global adjoint tomography example. Geophysical Journal International 219, 1225-1236.

SCEDC, 2013. Southern California Earthquake Center. Caltech, Dataset, doi:10.7909/C7903WD7903xH7901.

Tanaka, S., Siripunvaraporn, W., Boonchaisuk, S., Noisagool, S., Kim, T., Kawai, K., et al., 2019. Thai Seismic Array (TSAR) Project. 東京大学地震研究所彙報= Bulletin of the Earthquake Research Institute, University of Tokyo 94, 1-11.

Thurber, C., Zhang, H., Waldhauser, F., Hardebeck, J., Michael, A., Eberhart-Phillips, D., 2006. Three-dimensional compressional wavespeed model, earthquake relocations, and focal mechanisms for the Parkfield, California, region. Bulletin of the Seismological Society of America 96, S38-S49.

Tong, P., 2021a. Adjoint-state traveltime tomography for azimuthally anisotropic media and insight into the crustal structure of central California near Parkfield. Journal of Geophysical Research: Solid Earth 126, e2021JB022365.

Tong, P., 2021b. Adjoint-state traveltime tomography: Eikonal equation-based methods and application to the Anza area in Southern California. Journal of Geophysical Research: Solid Earth 126, e2021JB021818.

Tong, P., Li, T., Chen, J., Nagaso, M., 2023. Adjoint-state differential arrival time tomography. Geophysical Journal International, ggad416.

Tong, P., Yang, D., Huang, X., 2019. Multiple-grid model parametrization for seismic tomography with application to the San Jacinto fault zone. Geophysical Journal International 218, 200-223.

Townend, J., Zoback, M. D., 2004. Regional tectonic stress near the San Andreas fault in central and southern California. Geophysical Research Letters 31, L15S11.

VanDecar, J., Crosson, R., 1990. Determination of teleseismic relative phase arrival times using multi-channel cross-correlation and least squares. Bulletin of the Seismological Society of America 80, 150-169.

Vidale, J., 1988. Finite-difference calculation of travel times. Bulletin of the Seismological Society of America 78, 2062-2076.

Wu, S., Jiang, C., Schulte-Pelkum, V., Tong, P., 2022. Complex patterns of past and ongoing crustal deformations in Southern California revealed by seismic azimuthal anisotropy. Geophysical Research Letters 49, e2022GL100233.

Yang, T., Liu, F., Harmon, N., Le, K. P., Gu, S., Xue, M., 2015. Lithospheric structure beneath Indochina block from Rayleigh wave phase velocity tomography. Geophysical Journal International 200, 1582-1595.

Yu, C., Shi, X., Yang, X., Zhao, J., Chen, M., Tang, Q., 2017. Deep thermal structure of Southeast Asia constrained by S-velocity data. Marine Geophysical Research 38, 341-355.

Yu, Y., Gao, S. S., Liu, K. H., Yang, T., Xue, M., Le, K. P., 2017. Mantle transition zone discontinuities beneath the Indochina Peninsula: Implications for slab subduction and mantle upwelling. Geophysical Research Letters 44, 7159-7167.

Zhao, H., 2005. A fast sweeping method for Eikonal equations. Mathematics of Computation 74, 603-627.

Zhao, H., 2007. Parallel Implementations of the fast sweeping method. Journal of Computational Mathematics 25, 421-429.

# Appendix

## A. Derivation of adjoint sources

The core feature of the adjoint-state method is to solve the adjoint equation (5), in which the adjoint source takes the general form of

$$R_{n,m} = \left(\frac{\partial}{\partial T_n(x_{r,m})}\chi\right) = \frac{\partial \chi_t}{\partial T_n(x_{r,m})} + \frac{\partial \chi_{cs}}{\partial T_n(x_{r,m})} + \frac{\partial \chi_{cr}}{\partial T_n(x_{r,m})}. \tag{A1}$$

The three derivatives are detailed in Tong et al. (2023), which are also specified as follows

$$\frac{\partial \chi_t}{\partial T_n(x_{r,m})} = w_{n,m}\left(T_n(x_{r,m}) - T_n^{obs}(x_{r,m})\right), \tag{A2}$$

$$\begin{aligned}\frac{\partial \chi_{cs}}{\partial T_n(x_{r,m})} &= \sum_{i=1}^{N_r} w_{n,mi}\left(\left(T_n(x_{r,m}) - T_n(x_{r,i})\right) - \left(T_n^{obs}(x_{r,m}) - T_n^{obs}(x_{r,i})\right)\right) \\ &\quad - \sum_{j=1}^{N_r} w_{n,jm}\left(\left(T_n(x_{r,j}) - T_n(x_{r,m})\right) - \left(T_n^{obs}(x_{r,j}) - T_n^{obs}(x_{r,m})\right)\right) \\ &= 2\sum_{i=1}^{N_r} w_{n,mi}\left(\left(T_n(x_{r,m}) - T_n(x_{r,i})\right) \right. \\ &\quad \left. - \left(T_n^{obs}(x_{r,m}) - T_n^{obs}(x_{r,i})\right)\right), \end{aligned} \tag{A3}$$

$$\begin{aligned}\frac{\partial \chi_{cr}}{\partial T_n(x_{r,m})} &= \sum_{j=1}^{N_s} w_{nj,m}\left(\left(T_n(x_{r,m}) - T_j(x_{r,m})\right) - \left(T_n^{obs}(x_{r,m}) - T_j^{obs}(x_{r,m})\right)\right) \\ &\quad - \sum_{i=1}^{N_s} w_{in,m}\left(\left(T_i(x_{r,m}) - T_n(x_{r,m})\right) - \left(T_i^{obs}(x_{r,m}) - T_n^{obs}(x_{r,m})\right)\right) \\ &= 2\sum_{j=1}^{N_s} w_{nj,m}\left(\left(T_n(x_{r,m}) - T_j(x_{r,m})\right) \right. \\ &\quad \left. - \left(T_n^{obs}(x_{r,m}) - T_j^{obs}(x_{r,m})\right)\right), \end{aligned} \tag{A4}$$

## B. The sensitivity kernel with respect to earthquake location

By using the reciprocal principle, the synthetic traveltime $T_n(x_{r,m})$ from the $n$-th source $x_{s,n}$ and recorded by the $m$-th receiver $x_{r,m}$ equals the synthetic traveltime $\Gamma_m(x_{s,n})$ from the receiver $x_{r,m}$ to the source $x_{s,n}$ (refer to Tong et al. (2023) for detailed discussion). Here the function $\Gamma_m(x)$ describes the wavefront traveltime from the $m$-th receiver $x_{r,m}$ to any position $x$. Consequently, the objective functions (2)-(4) can be reformulated as

$$\chi_t = \sum_{n=1}^{N_s} \sum_{m=1}^{N_r} \frac{w_{n,m}}{2} \left( \Gamma_m(x_{s,n}) + \tau_n - T_n^{obs}(x_{s,m}) \right)^2, \tag{B1}$$

$$\chi_{cs} = \sum_{n=1}^{N_s} \sum_{m=1}^{N_r} \sum_{i=1}^{N_r} \frac{w_{n,mi}}{2} \left( \left( \Gamma_m(x_{s,n}) - \Gamma_i(x_{s,n}) \right) - \left( T_n^{obs}(x_{r,m}) - T_n^{obs}(x_{r,i}) \right) \right)^2, \tag{B2}$$

$$\chi_{cr} = \sum_{n=1}^{N_s} \sum_{j=1}^{N_s} \sum_{m=1}^{N_r} \frac{w_{nj,m}}{2} \left( \left( \Gamma_m(x_{s,n}) - \Gamma_m(x_{s,j}) \right) + \tau_n - \tau_j - \left( T_n^{obs}(x_{r,m}) - T_j^{obs}(x_{r,m}) \right) \right)^2. \tag{B3}$$

The sensitivity kernels of the objective function with respect to $x_{s,n}$ are derived as follows

$$\nabla_{x_{s,n}} \chi = \nabla_{x_{s,n}} \chi_t + \nabla_{x_{s,n}} \chi_{cs} + \nabla_{x_{s,n}} \chi_{cr}, \tag{B4}$$

in which

$$\nabla_{x_{s,n}} \chi_t = \sum_{m=1}^{N_r} w_{k,m} \left( \Gamma_m(x_{s,n}) + \tau_n - T_n^{obs}(x_{r,m}) \right) \nabla \Gamma_m(x_{s,n}), \tag{B5}$$

$$\nabla_{x_{s,n}} \chi_{cs} = \sum_{m=1}^{N_r} \sum_{i=1}^{N_r} w_{n,mi} \left( \left( \Gamma_m(x_{s,n}) - \Gamma_i(x_{s,k}) \right) - \left( T_n^{obs}(x_{r,m}) - T_n^{obs}(x_{r,i}) \right) \right) \left( \nabla \Gamma_m(x_{s,n}) - \nabla \Gamma_i(x_{s,n}) \right)$$

$$= 2 \sum_{m=1}^{N_r} \sum_{i=1}^{N_r} w_{n,mi} \left( \left( \Gamma_m(x_{s,n}) - \Gamma_i(x_{s,n}) \right) - \left( T_n^{obs}(x_{r,m}) - T_n^{obs}(x_{r,i}) \right) \right) \nabla \Gamma_m(xx_{s,n}), \tag{B6}$$

$$\nabla_{x_{s,n}} \chi_{cr} = \sum_{j=1}^{N_s} \sum_{m=1}^{N_r} w_{nj,m} \left( \left( \Gamma_m(x_{s,n}) - \Gamma_m(x_{s,j}) \right) + \tau_n - \tau_j - \left( T_n^{obs}(x_{r,m}) - T_j^{obs}(x_{r,m}) \right) \right) \nabla \Gamma_m(x_{s,n})$$

$$- \sum_{i=1}^{N_s} \sum_{m=1}^{N_r} w_{in,m} \left( \left( \Gamma_m(x_{s,i}) - \Gamma_m(x_{s,n}) \right) + \tau_i - \tau_n - \left( T_i^{obs}(x_{r,m}) - T_n^{obs}(x_{r,m}) \right) \right) \nabla \Gamma_m(x_{s,n})$$

$$= 2 \sum_{j=1}^{N_s} \sum_{m=1}^{N_r} w_{nj,m} \left( \left( \Gamma_m(x_{s,n}) - \Gamma_m(x_{s,j}) \right) + \tau_n - \tau_j - \left( T_n^{obs}(x_{r,m}) - T_j^{obs}(x_{r,m}) \right) \right) \nabla \Gamma_m(x_{s,n}). \tag{B7}$$

Similarly, the sensitivity kernels of the objective function with respect to $\tau_n$ are derived as follows

$$\partial_{\tau_n} \chi = \partial_{\tau_n} \chi_t + \partial_{\tau_n} \chi_{cs} + \partial_{\tau_n} \chi_{cr}, \tag{B8}$$

in which

$$\partial_{\tau_n} \chi_t = \sum_{m=1}^{N_r} w_{n,m} \left( \Gamma_m(x_{s,n}) + \tau_n - T_n^{obs}(x_{r,m}) \right), \tag{B9}$$

$$\partial_{\tau_n} \chi_{cs} = 0 \tag{B10}$$

$$\partial_{\tau_n}\chi_{cr} = \sum_{j=1}^{N_s}\sum_{m=1}^{N_r} w_{nj,m}\left(\left(\Gamma_m(x_{s,n}) - \Gamma_m(x_{s,j})\right) + \tau_n - \tau_j - \left(T_n^{obs}(x_{r,m}) - T_j^{obs}(x_{r,m})\right)\right)$$

$$- \sum_{i=1}^{N_s}\sum_{m=1}^{N_r} w_{in,m}\left(\left(\Gamma_m(x_{s,i}) - \Gamma_m(x_{s,n})\right) + \tau_i - \tau_n - \left(T_i^{obs}(x_{r,m}) - T_n^{obs}(x_{r,m})\right)\right)$$

$$= 2\sum_{j=1}^{N_s}\sum_{m=1}^{N_r} w_{nj,m}\left(\left(\Gamma_m(x_{s,n}) - \Gamma_m(x_{s,j})\right) + \tau_n - \tau_j - \left(T_n^{obs}(x_{r,m}) - T_j^{obs}(x_{r,m})\right)\right). \quad (B11)$$

List of Figures

1. Figure 1: The workflow of TomoATT to implement adjoint-state traveltime tomography, including anisotropic velocity model determination and earthquake location.

2. Figure 2: Validation of the Eikonal solver for calculating traveltime. (a) The true traveltime field $T_{true}(x)$ calculated in the velocity model. The red star indicates the earthquake. The black curves represent traveltime isochrones at an interval of 20 s. (b) Traveltime errors at surface stations (depth = 0 km) relative to epicenter distance. The blue, yellow, green, and red curves show numerical errors with grid spacings of 20, 10, 5, and 2.5 km, respectively. Oscillations in the blue curve may arise from interpolation errors on the sparse grid. (c) The left column shows the calculated traveltime fields $T_{cal}(x)$ with grid spacings of 20, 10, 5, and 2.5 km. True and calculated traveltime isochrones are denoted by black and white dashed curves, respectively. The right column displays the numerical error fields $T_{cal}(x) - T_{true}(x)$ for each grid spacing, with the $L_1$ norm errors quantified.

3. Figure 3: Validation of the Eikonal solver for teleseismic differential arrival time. (a) The teleseismic earthquake (red star) and the study region (red box). Black dashed circles denote epicenter distances of 30° and 60°. (b) Calculated traveltime field within the study region. The blue triangles denote stations deployed on the surface. (c) Histogram of differential arrival misfits $\Delta T_{cal} - \Delta T_{TauP}$. The mean value and standard deviation are $-0.021$ s and $0.03$ s, respectively, verifying the accuracy of the Eikonal solver.

4. Figure 4: A toy experiment illustrating the multipathing phenomenon. (a) P-wave velocity model. The blue triangles denote stations, and the red star marks the earthquake. White curves represent traveltime isochrones at an interval of 1 s. (b) Sensitivity kernel computed with artificial adjoint sources $R_{n,m} = 1$. The adjoint-state method captures the sensitivity along multiple paths, demonstrating its capability to address the multipathing phenomenon.

5. Figure 5: A toy experiment demonstrating multiple-grid parameterization and kernel density normalization in tomography. (a) Initial velocity model with 8 stations on the surface (blue triangles) and 10,000 unevenly distributed earthquakes. (b) Velocity perturbation of the target model relative to the initial model. (c) Original sensitivity kernel at the first iteration, rescaled to $[-1,1]$. (d) Imaging result after 40 iterations using the original sensitivity kernel. (e)-(f) Sensitivity kernel and imaging result after applying multiple-grid parameterization. Dots of the same color represents nodes in the same inversion grid. (g)-(l) Sensitivity kernels and imaging results after applying both multiple-grid parameterization and kernel density regularization with different coefficients $\zeta = 0.3, 0.6, 0.9$.

6. Figure 6: Illustration of multi-level parallelization in TomoATT. (a) Level 1: Source parallelization demonstrates numerically solving Eikonal equations in parallel for two earthquakes. The blue triangles denote stations, and the red star marks the earthquake. Gray circles denote grid nodes discretizing the computational domain. Level 2: A 2-by-2 domain decomposition. Arrows indicate communications between processors managing adjacent subdomains. Level 3: Hyperplane stepping parallelization. All grid nodes within a subdomain are categorized into multiple hyperplanes (oblique lines in different colors). The grid nodes on the same hyperplane (circles of the same color) can be processed in parallel. (b) Log-log plots showing the speed-up and memory usage relative the number of processors for each parallelization method.

7. Figure 7: Model setting of a synthetic experiment. (a) Velocity perturbation of the true model relative to the initial model. Yellow bars represent azimuthal anisotropy, aligned with fast velocity directions. Blue triangles are stations on the surface. True earthquake hypocenters are evenly distributed at depths of 10 km (red dots), 20 km (green dots), and 30 km (black dots). (b) Initial earthquake hypocenters. (c) Histograms of horizontal offsets of hypocenters, vertical offsets of hypocenters, and origin time offsets.

8. Figure 8: Figure 8. Test results in the synthetic experiment. Notations follow Figure 7. (a) Test 1: Earthquake locations determined in the true model, verifying the relocation function. (b) Test 2: Earthquake locations determined in the initial model, showing slight deviation due to inaccurate velocity and anisotropy. (c) Test 3: Velocity and anisotropic parameters updated with the true earthquake locations, verifying the tomography function (d) Test 4: Velocity and anisotropic parameters updated with the initial earthquake location, resulting in noticeable distortions and artifacts due to source uncertainty. (e) Test 5: Simultaneous inversion of model parameters and earthquake locations.

9. Figure 9: Tectonic setting and imaging results in central California near Parkfield. (a) Topography map. Red box outlines the study region. Earthquakes and stations are denoted by red dots and blue triangles, respectively. Black lines mark active faults. (b) Horizontal sections of P-wave velocity perturbation relative to the horizontal average. Coastline and the San Andreas Fault (SAF) are indicated by thinner and thicker black lines, respectively. Key tectonic features are labeled: FT (Franciscan Terrane), ST (Salinian Terrane), SMB (Santa Maria Basin), and TR (Transverse Ranges). (c) Horizontal sections of azimuthal anisotropy. Yellow bars align with the fast velocity directions. Black, red, and blue lines denote the creeping, transitional, and locked segments of the SAF.

10. Figure 10: Tectonic setting and imaging results in Thailand and adjacent regions. (a) Topography map. Black dashed lines denote major faults: DBPF (Dien Bien Phu Fault), WCF (Wang-Chao Fault), and TPF (Three Pagodas Fault). The solid black line indicates the Khorat Plateau. Stations are denoted by blue triangles. (b) Distribution of teleseismic earthquakes used for the inversion. Dashed circles represent epicenter distances of **30°** and **60°**. (c) Horizontal and vertical sections of velocity perturbation relative to the horizontal average. Vertical section locations are plotted as green dashed lines. KP represents the Khorat Plateau. In **BB′** profile, two black arrows indicate possible pathways of mantle upwelling, shown as low-velocity perturbations extending from the surface down to the upper mantle.

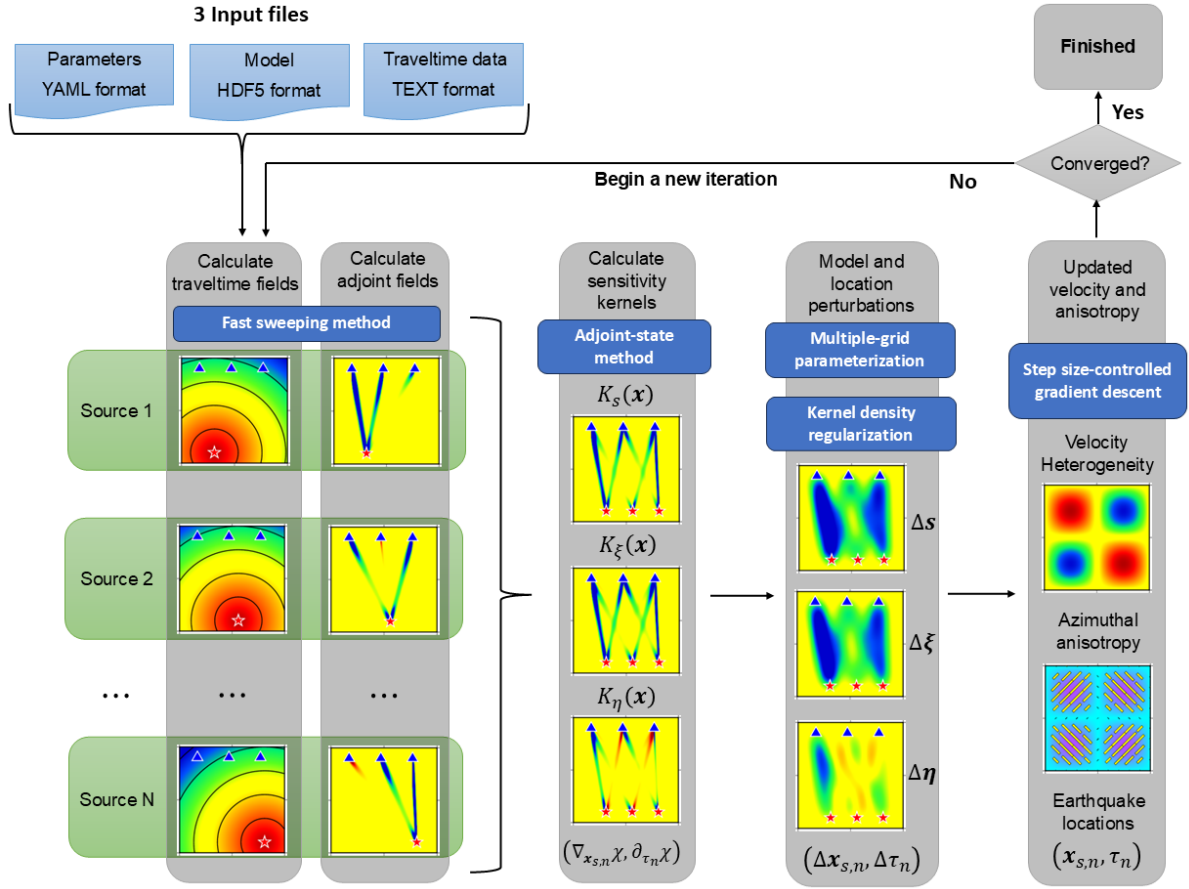

**Figure 1.** The workflow of TomoATT to implement adjoint-state traveltime tomography, including anisotropic velocity model determination and earthquake location.

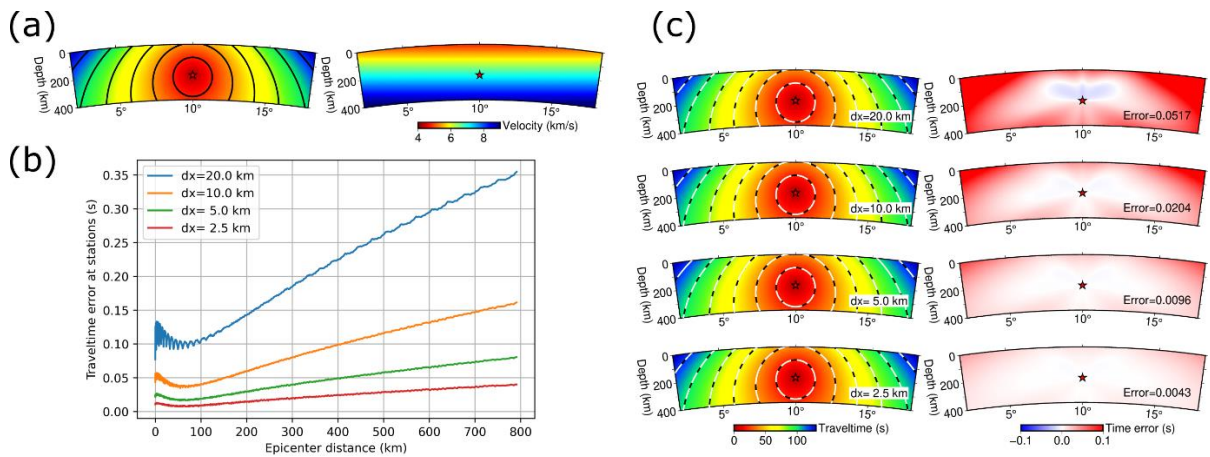

**Figure 2.** Validation of the Eikonal solver for calculating traveltime. (a) The true traveltime field $T_{true}(x)$ calculated in the velocity model. The red star indicates the earthquake. The black curves represent traveltime isochrones at an interval of 20 s. (b) Traveltime errors at surface stations (depth = 0 km) relative to epicenter distance. The blue, yellow, green, and red curves show numerical errors with grid spacings of 20, 10, 5, and 2.5 km, respectively. Oscillations in the blue curve may arise from interpolation errors on the sparse grid. (c) The left column shows the calculated traveltime fields $T_{cal}(x)$ with grid spacings of 20, 10, 5, and 2.5 km. True and calculated traveltime isochrones are denoted by black and white dashed curves, respectively. The right column displays the numerical error fields $T_{cal}(x) - T_{true}(x)$ for each grid spacing, with the $L_1$ norm errors quantified.

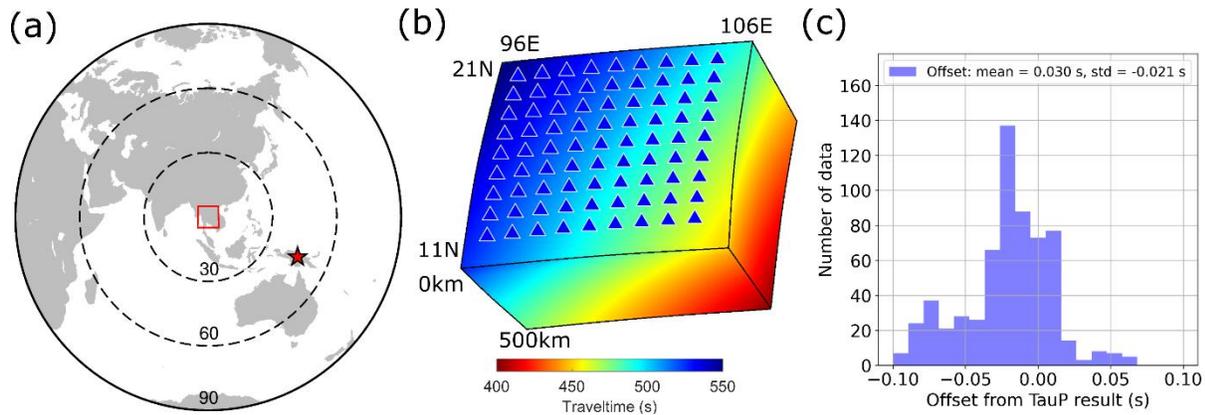

**Figure 3.** Validation of the Eikonal solver for teleseismic differential arrival time. (a) The teleseismic earthquake (red star) and the study region (red box). Black dashed circles denote epicenter distances of 30° and 60°. (b) Calculated traveltime field within the study region. The blue triangles denote stations deployed on the surface. (c) Histogram of differential arrival misfits $\Delta T_{cal} - \Delta T_{TauP}$. The mean value and standard deviation are $-0.021$ s and 0.03 s, respectively, verifying the accuracy of the Eikonal solver.

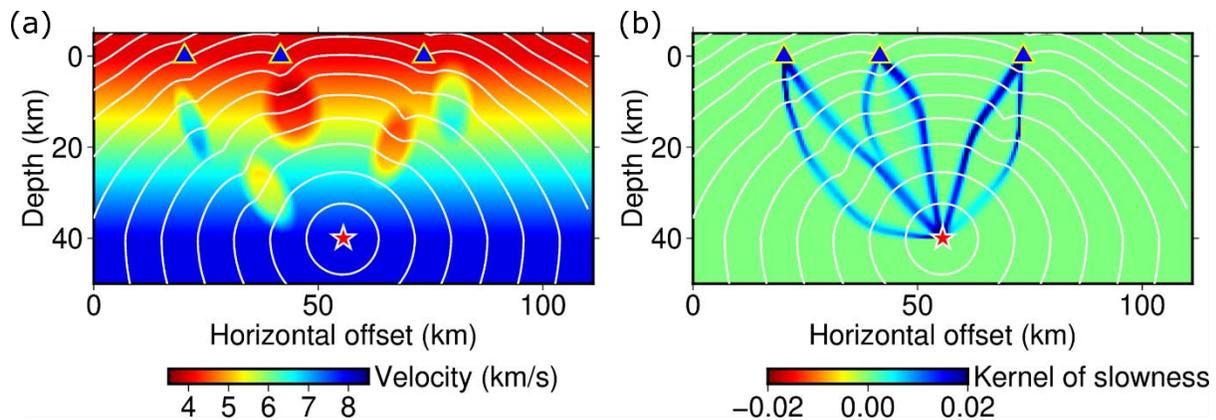

**Figure 4.** A toy experiment illustrating the multipathing phenomenon. (a) P-wave velocity model. The blue triangles denote stations, and the red star marks the earthquake. White curves represent traveltime isochrones at an interval of 1 s. (b) Sensitivity kernel computed with artificial adjoint sources $R_{n,m} = 1$. The adjoint-state method captures the sensitivity along multiple paths, demonstrating its capability to address the multipathing phenomenon.

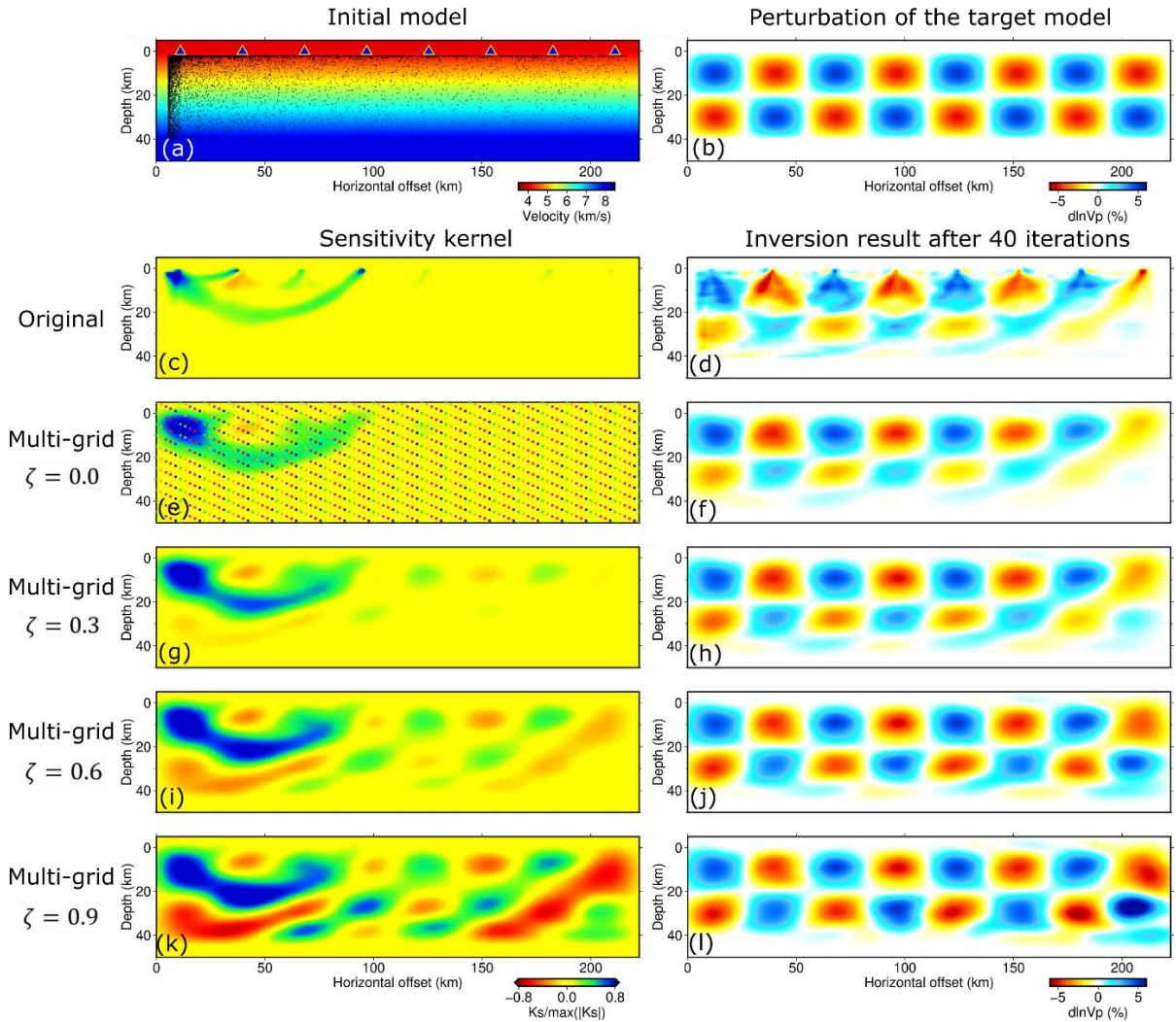

**Figure 5.** A toy experiment demonstrating multiple-grid parameterization and kernel density normalization in tomography. (a) Initial velocity model with 8 stations on the surface (blue triangles) and 10,000 unevenly distributed earthquakes. (b) Velocity perturbation of the target model relative to the initial model. (c) Original sensitivity kernel at the first iteration, rescaled to $[-1,1]$. (d) Imaging result after 40 iterations using the original sensitivity kernel. (e)-(f) Sensitivity kernel and imaging result after applying multiple-grid parameterization. Dots of the same color

represents nodes in the same inversion grid. (g)-(l) Sensitivity kernels and imaging results after applying both multiple-grid parameterization and kernel density regularization with different coefficients $\zeta = 0.3, 0.6, 0.9$.

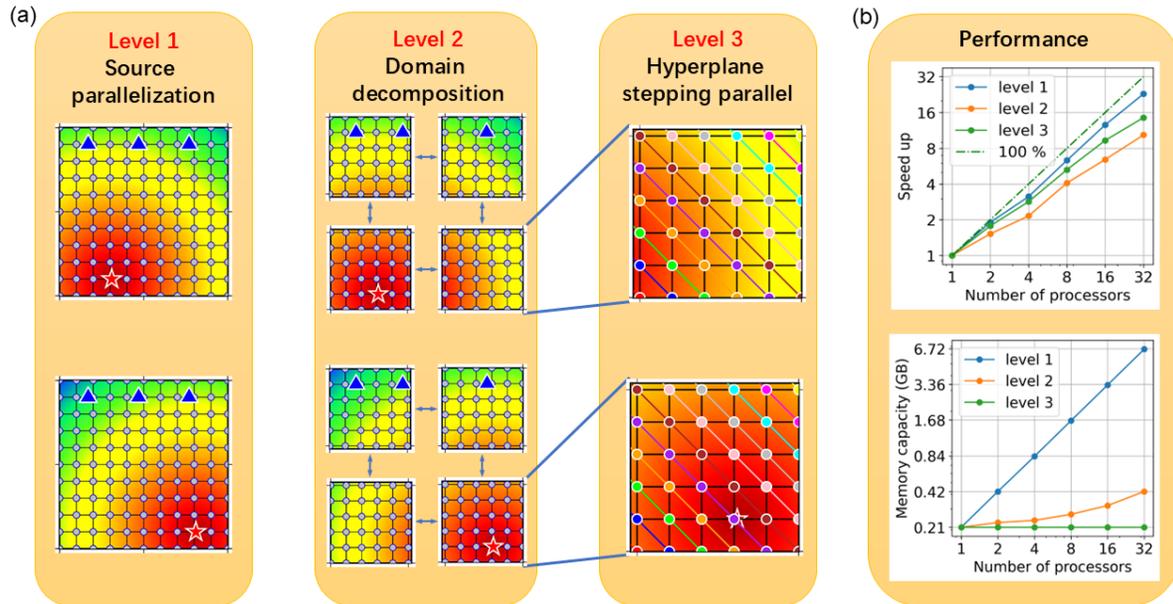

**Figure 6.** Illustration of multi-level parallelization in TomoATT. (a) Level 1: Source parallelization demonstrates numerically solving Eikonal equations in parallel for two earthquakes. The blue triangles denote stations, and the red star marks the earthquake. Gray circles denote grid nodes discretizing the computational domain. Level 2: A 2-by-2 domain decomposition. Arrows indicate communications between processors managing adjacent subdomains. Level 3: Hyperplane stepping parallelization. All grid nodes within a subdomain are categorized into multiple hyperplanes (oblique lines in different colors). The grid nodes on the same hyperplane (circles of the same color) can be processed in parallel. (b) Log-log plots showing the speed-up and memory usage relative the number of processors for each parallelization method.

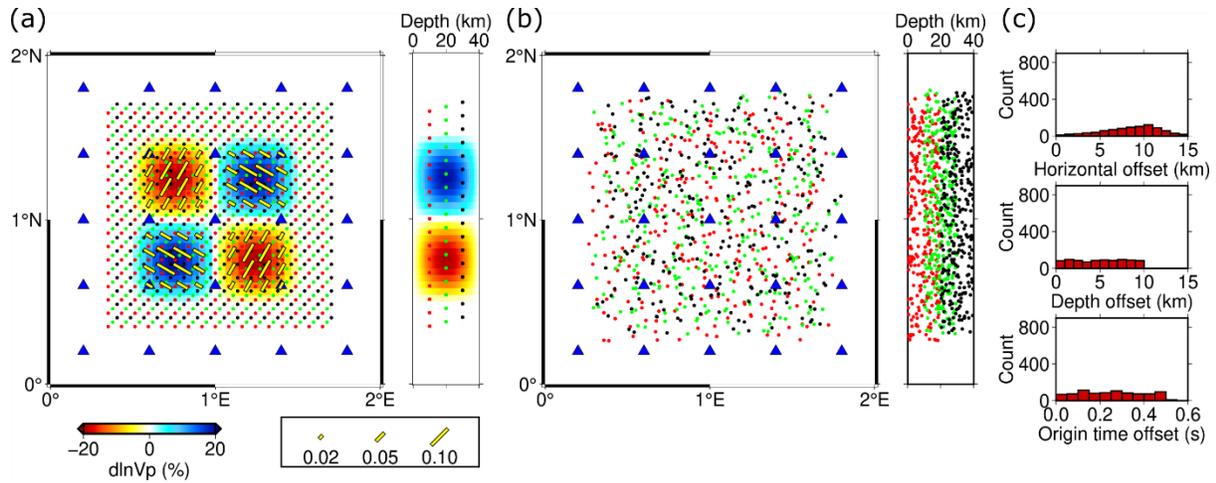

**Figure 7.** Model setting of a syntehtic experiment. (a) Velocity perturbation of the true model relative to the initial model. Yellow bars represent azimuthal anisotropy, aligned with fast velocity directions. Blue triangles are stations on the surface. True earthquake hypocenters are evenly distributed at depths of 10 km (red dots), 20 km (green dots), and 30 km (black dots). (b) Initial earthquake hypocenters. (c) Histograms of horizontal offsets of hypocenters, vertical offsets of hypocenters, and origin time offsets.

**Figure 8.** Test results in the synthetic experiment. Notations follow Figure 7. (a) Test 1: Earthquake locations determined in the true model, verifying the relocation function. (b) Test 2: Earthquake locations determined in the initial model, showing slight deviation due to inaccurate velocity and anisotropy. (c) Test 3: Velocity and anisotropic parameters updated with the true earthquake locations, verifying the tomography function (d) Test 4: Velocity and anisotropic parameters updated with the initial earthquake location, resulting in noticeable distortions and artifacts due to source uncertainty. (e) Test 5: Simultaneous inversion of model parameters and earthquake locations.

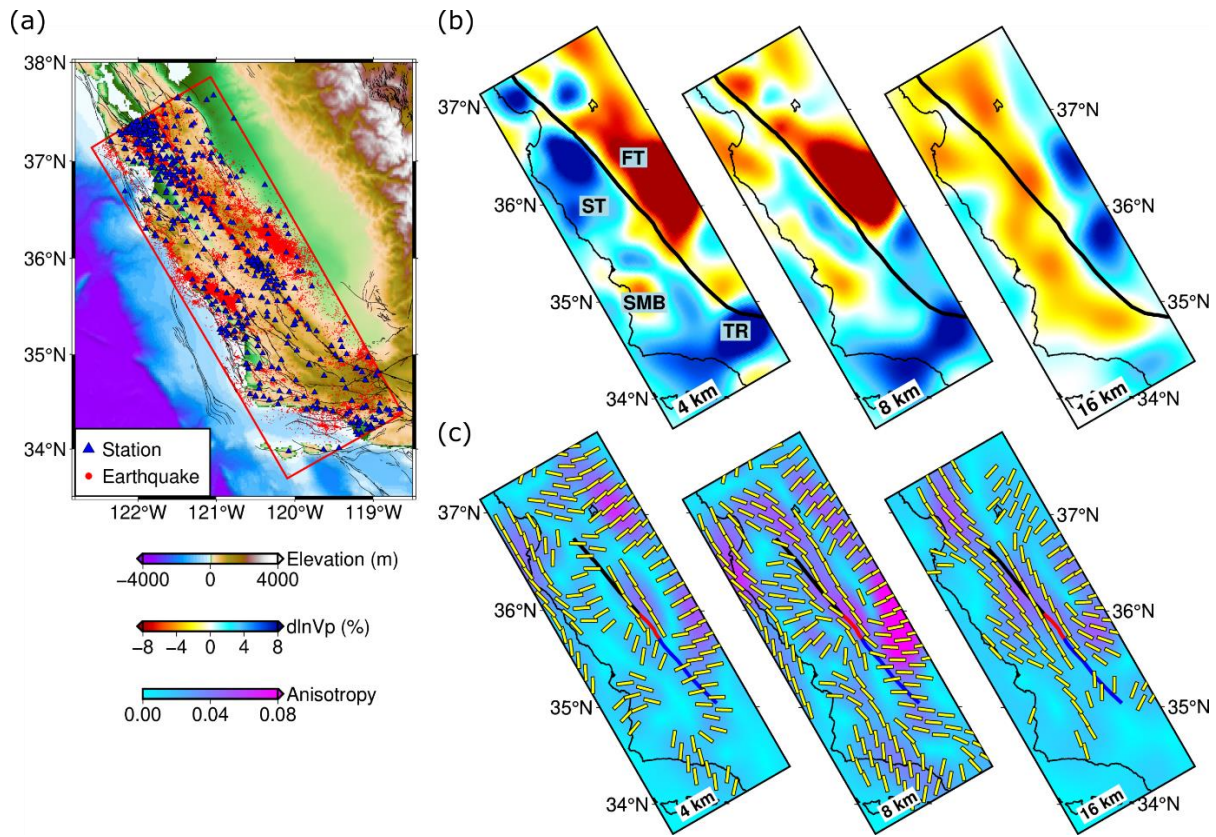

**Figure 9.** Tectonic setting and imaging results in central California near Parkfield. (a) Topography map. Red box outlines the study region. Earthquakes and stations are denoted by red dots and blue triangles, respectively. Black lines mark active faults. (b) Horizontal sections of P-wave velocity perturbation relative to the horizontal average. Coastline and the San Andreas Fault (SAF) are indicated by thinner and thicker black lines, respectively. Key tectonic features are labeled: FT (Franciscan Terrane), ST (Salinian Terrane), SMB (Santa Maria Basin), and TR (Transverse Ranges). (c) Horizontal sections of azimuthal anisotropy. Yellow bars align with the fast velocity directions. Black, red, and blue lines denote the creeping, transitional, and locked segments of the SAF.

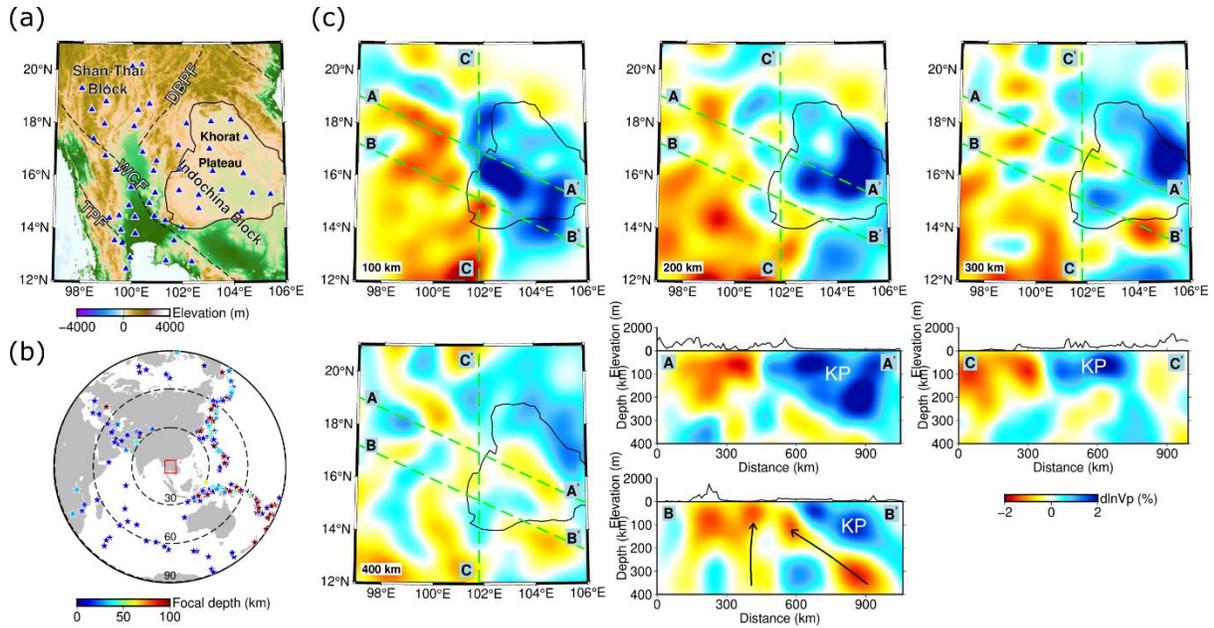

**Figure 10.** Tectonic setting and imaging results in Thailand and adjacent regions. (a) Topography map. Black dashed lines denote major faults: DBPF (Dien Bien Phu Fault), WCF (Wang-Chao Fault), and TPF (Three Pagodas Fault). The solid black line indicates the Khorat Plateau. Stations are denoted by blue triangles. (b) Distribution of teleseismic earthquakes used for the inversion. Dashed circles represent epicenter distances of **30°** and **60°**. (c) Horizontal and vertical sections of velocity perturbation relative to the horizontal average. Vertical section locations are plotted as green dashed lines. KP represents the Khorat Plateau. In **BB′** profile, two black arrows indicate possible pathways of mantle upwelling, shown as low-velocity perturbations extending from the surface down to the upper mantle.

**Table 1.**

The computational time and memory usage for each parallelization method.

| Number of processors | Source parallelization | Domain decomposition | Hyperplane stepping parallelization |
|---|---|---|---|
| 1 | 662.62 s / 0.21 GB | - | - |
| 2 | 348.03 s / 0.42 GB | 435.17 s / 0.23 GB | 372.72 s / 0.21 GB |
| 4 | 211.41 s / 0.83 GB | 307.20 s / 0.24 GB | 233.11 s / 0.21 GB |
| 8 | 104.62 s / 1.66 GB | 162.59 s / 0.27 GB | 125.53 s / 0.21 GB |
| 16 | 52.64 s / 3.32 GB | 102.99 s / 0.32 GB | 71.14 s / 0.21 GB |
| 32 | 28.85 s / 6.65 GB | 63.99 s / 0.42 GB | 45.73 s / 0.21 GB |